\documentclass[12pt]{article}
\usepackage{hyperref}
\usepackage{natbib}
\usepackage{authblk}
\usepackage{booktabs}
\usepackage{url}
\usepackage{verbatim}
\usepackage{multirow}
\usepackage{setspace}
\usepackage{graphicx}
\usepackage[centertags]{amsmath}
\usepackage{amsfonts}
\usepackage{multirow}
\usepackage{amssymb}
\usepackage{amsmath}
\usepackage{amsthm}
\usepackage[noend]{algpseudocode}
\usepackage{algorithm}
\usepackage[top=1.5in, bottom=1.5in, left=1in, right=1in]{geometry}
\theoremstyle{plain}
\newtheorem{thm}{Theorem}
\newtheorem{pro}{Property}

\theoremstyle{definition}

\theoremstyle{remark}
\newtheorem{rem}{Remark}[section]

\newcounter{eqn}

\makeatletter
\newcommand{\putindeepbox}[2][0.7\baselineskip]{{%
    \setbox0=\hbox{#2}%
    \setbox0=\vbox{\noindent\hsize=\wd0\unhbox0}
    \@tempdima=\dp0
    \advance\@tempdima by \ht0
    \advance\@tempdima by -#1\relax
    \dp0=\@tempdima
    \ht0=#1\relax
    \box0
}}
\makeatother

\usepackage{setspace}

\title{co-BPM: a Bayesian Model for Divergence Estimation}
\author[1]{Kun Yang}
\affil[1]{Institute of Computational and Mathematical Engineering, Stanford University}
\author[2]{Hao Su}
\affil[2]{Department of Computer Science, Stanford University}
\author[3]{Wing Hung Wong}
\affil[3]{Department of Statistics, Stanford University}
\affil[3]{Department of Health Research and Policy, Stanford University}
\date{}

\begin{document}

\maketitle

\begin{abstract}
Divergence is not only an important mathematical concept in information theory, but also applied to machine learning problems such as low-dimensional embedding, manifold learning, clustering, classification, and anomaly detection. We proposed a bayesian model---co-BPM---to characterize the discrepancy of two sample sets, i.e., to estimate the divergence of their underlying distributions. In order to avoid the pitfalls of plug-in methods that estimate each density independently, our bayesian model attempts to learn a coupled binary partition of the sample space that best captures the landscapes of both distributions, then make direct inference on their divergences. The prior is constructed by leveraging the sequential buildup of the coupled binary partitions and the posterior is sampled via our specialized MCMC. Our model provides a unified way to estimate various types of divergences and enjoys convincing accuracy. We demonstrate its effectiveness through simulations, comparisons with the \emph{state-of-the-art} and a real data example.

\end{abstract}

\section{Introduction}
Divergence between two distributions is of significant importance in various disciplines such as statistics, information theory and machine learning. In statistics, hypothesis testing procedures such as Kolmogorov-Simirnov test and Cramer-von Mises test are based on some discrepancy measurements of the empirical distributions, and a common goal in bayesian experimental design \citep{Chaloner1995} is to maximize the expected Kullback-Leibler (KL) divergence between the prior and the posterior. In information theory, theoretic quantities such as mutual information and Shannon Entropy are derived from divergences. In machine learning, different divergences such as Total Variation, Hellinger Distance and KL divergence are often applied as a dissimilarity measure in manifold learning \citep{Donoho2003}, classification and anomaly detection, etc. Recently, they are also applied to developing efficient methods for a robust empirical risk minimization problem \citep{Namkoong2016}.

Given two probability measures $P_1$ and $P_2$ defined on domain $\Omega$, we consider the case that $P_1$ and $P_2$ have their densities $p_1$ and $p_2$, a general form of divergence is given by
\[D_\phi(p_1, p_2) = \int p_1(x)\phi\Big(\frac{p_2(x)}{p_1(x)}\Big)dx\]
where $\phi$ is a convex function. $D_\phi$ is the class of Ali-Silvey distances \citep{Ali1966}, also known as $f-$divergences. Many common divergences, such as KL divergence ($\phi(x) = -\log(x)$), $\alpha-$divergence (e.g., $\phi(x) = 4 / (1 - \alpha^2)\cdot(1 - x^{(\alpha + 2) / 2})$ when $\alpha \neq \pm 1$), Hellinger distance ($\sqrt{D_\phi}$ and $\phi(x) = 1 - \sqrt{x}$), and total variation distance ($\phi(x) = |x - 1|$), are special cases of $f-$divergence, coinciding with a particular choice of $\phi$.

We are interested in estimating $D_\phi$ from two sample sets $\mathcal{X}$ and $\mathcal{Y}$, where $\mathcal{X} = \{x_i\}_{i = 1}^{n_1}$  and $\mathcal{Y} = \{y_i\}_{i = 1}^{n_2}$ are iid samples drawn from $P_1$ and $P_2$ respectively. A conceptually simple approach proceeds in two-steps: estimate the densities $\hat{p}_1$ and $\hat{p}_2$ with $\mathcal{X}$ and $\mathcal{Y}$ independently then calculate $D_\phi(\hat{p}_1, \hat{p}_2)$. However, this approach is unattractive for the following reasons: 1) multivariate density estimation itself is a challenging problem and often more difficult than comparing distributions, poorly estimated densities lead to high variance and bias in subsequent divergence estimation \citep{Ma2011}; 2) with $\hat{p}_1$ and $\hat{p}_2$ estimated independently, we are often unable to compute their divergence $D_\phi(\hat{p}_1, \hat{p}_2)$ analytically, and the application of numeric or approximation methods introduces additional error terms \citep{Sugiyama2012}. Hence, throughout this paper, we focus on the class of single-shot methods, i.e., comparing sample sets directly and simultaneously.

Divergence estimation, especially KL divergence, has a long history and rich literature, we review several methods in the context of information theory and machine learning. Interested readers may refer to \citep{Sricharan2010, Wang2009, Leonenko2008, Poczos2011, Ma2011, Poczos2012} for other recent developments. \citep{Wang2005} proposes a domain-partitioning method that constructs an adaptive partition of the sample space with respect to $\mathcal{Y}$, then Radon-Nikodym derivative $p_1(x) / p_2(x)$ can be estimated with empirical probability mass (with the correction term on the boundary) in each sub-region. In 1-dimension, strong consistency is established and several algorithmic refinements are suggested. In multi-dimensions, a partitioning scheme is also mentioned based on the heuristics in 1-dimensional case, however, there are not enough numeric simulations to justify the heuristics. \citep{Nguyen2010} derives a variational characterization of the $f-$divergence in terms of a bayes decision problem, which is exploited to develop an $M-$estimator. The theoretical results of consistency and convergence rates are provided. It is also shown that the estimation procedure can be cast into a finite-dimensional convex program and solved efficiently when the functional class is defined by reproducing kernel Hilbert spaces. \citep{Sugiyama2012} introduces the least-squares density-difference estimator developed under the framework of kernel regularized least-squares estimation. The finite sample error bound shows that it achieves the optimal convergence rate. They also demonstrate several pattern recognition applications.

Despite of the success of these methods, they are insufficient in several scenarios. Firstly, since $\mathcal{X}$ and $\mathcal{Y}$ are two sets of random samples, one may be interested in the confidence (credible) intervals of estimated divergences \citep{Sricharan2010}, but deriving such quantities are non-trivial for these methods. Secondly, these methods and theories are all set up for a specific type of divergence and are usually not directly applicable to others. Thirdly, based on our practical experience, previous work \citep{Sricharan2010, Nguyen2010, Perez-Cruz2008} faces challenges to scale to higher dimensions, even though most of the methods claim that there are no technical difficulties as the number of dimensions increases.

In this paper, we introduce a single-shot bayesian model, which we name as {\bf co}upled {\bf B}inary {\bf P}artition {\bf M}odel (co-BPM). Our model is inspired by two key ideas. Firstly, we use an adaptive partitioning scheme, which has been demonstrated to be more scalable (see Table \ref{extreme}) than traditional domain-partitioning methods such as histogram \citep{Wang2005} or regular paving~\citep{Sainudiin2013}. Secondly, instead of partitioning based upon $\mathcal{Y}$ alone as in~\citep{Wang2005}, we force the domain $\Omega$ to be \emph{coupled} so that it is partitioned with respect to $\mathcal{X}$ and $\mathcal{Y}$ simultaneously. Therefore, our model is capable of capturing the landscapes of both $\mathcal{X}$ and $\mathcal{Y}$, thus is more effective in estimating their divergences.

We highlight our contributions as follows:
\begin{enumerate}
\item co-BPM is a \emph{single-shot domain-partition} based bayesian model for divergence estimation, that is scalable in both dimension and sample size.
\item Our specifically tailored MCMC sampling algorithm exploits the sequential buildup of binary partition for rapid mixing.
\item co-BPM enjoys convincing accuracy and demonstrates superior performance in machine learning tasks such as clustering and classification.
\end{enumerate}

We validate our model by several examples: 1) two sanity tests to demonstrate that our model is sensitive to the differences among samples; 2) 1-dim and 3-dim numeric examples to assess the estimation accuracies; 3) simulations in 3-dim used to compare with the \emph{state-of-the-art} such as \citep{Nguyen2010} and \citep{Perez-Cruz2008}; 4) real data applications to image clustering and classification.

The rest of the paper is organized as follows: Section 2 introduces some notations used throughout our discussion and preliminaries that motivate our model; Section 3 introduces the Coupled Binary Partition Model and the specially tailored MCMC algorithm to sample posterior effectively; extensive simulations and comparisons are presented in Section 4; we draw the conclusions in Section 5.

\begin{figure}
  \center
  \includegraphics[width = 0.8\textwidth]{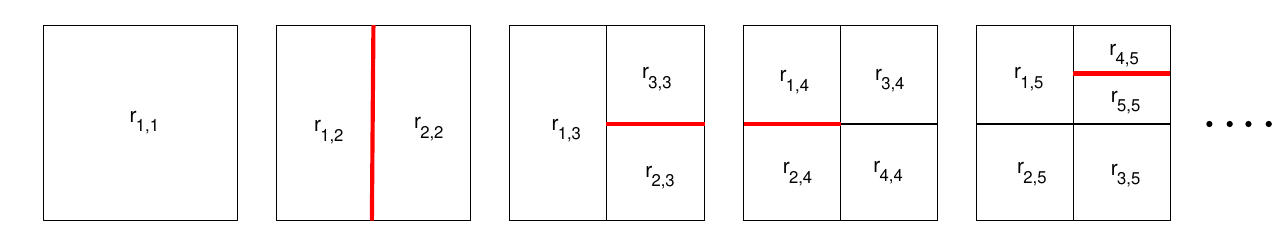}
  \caption{A \textbf{binary partition sequence} demonstrates its sequential structure. Each red line indicates the decision to halve the sub-region alone it.}
  \label{fig:sbp}
\end{figure}

\begin{figure}
\center
  \includegraphics[width = .4\textwidth]{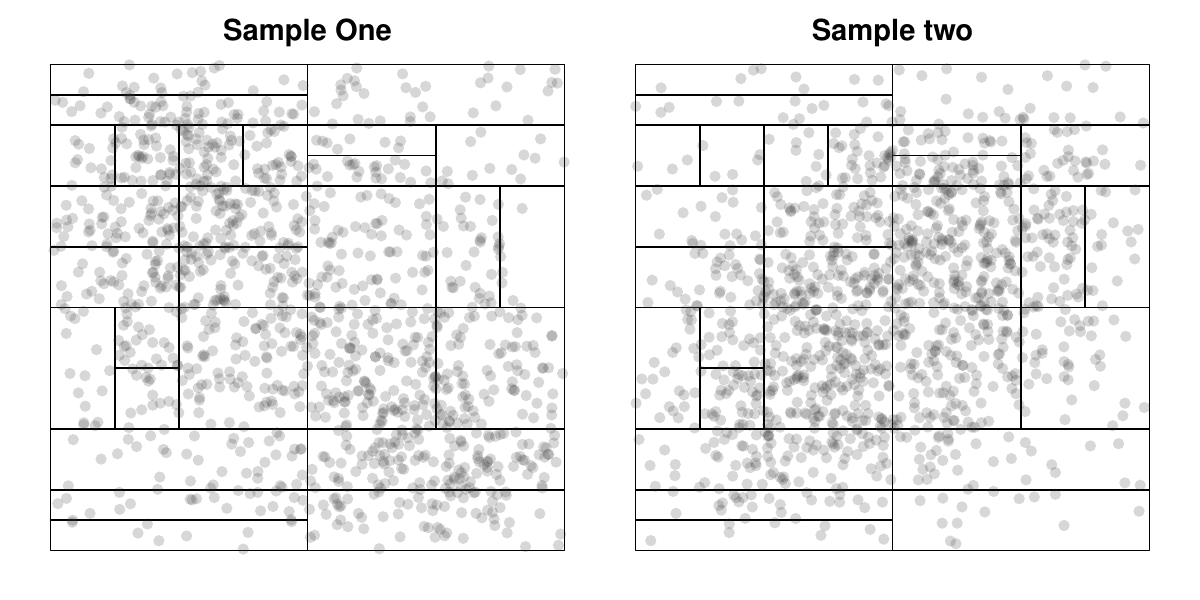}
  \caption{An illustration of \textbf{coupled binary partition} of two samples through the coupled binary partition model. Two samples are drawn from two different distributions and their sampling spaces are partitioned in tandem.}
  \label{joint_par}
\end{figure}

\section{Notation and Preliminaries}
Without loss of generality, we assume the domain $\Omega = [0, 1]^d$ with dimension $d$. A coordinate-wise binary partition has hierarchical structure that is constructed sequentially (Figure \ref{fig:sbp} gives an illustration of the sequence in $[0, 1]^2$): starting with $\mathcal{B}_1 = \{r_{1, 1} = \Omega\}$ at depth $1$, action $a_1$ is taken to split $\Omega$ into $\mathcal{B}_2 = \{r_{1, 2}, r_{2, 2}\}$ along the middle of some coordinate; then at depth 2, action $a_2$ is taken to halve some sub-region in $\mathcal{B}_2$ into $\mathcal{B}_3 = \{r_{1, 3}, r_{2, 3}, r_{3, 3}\}$ evenly. The process keeps on till the specified depth is reached. Given the maximum depth $l$, the decision sequence is denoted as $A_l = (a_1, ..., a_{l - 1})$. $A_l$ uniquely determines a partition of $\Omega$, whose sub-regions are denoted as $\{r_{1, l}, ..., r_{l, l}\}$. It is necessary to point out that different decision sequences may result in the same partition. Figure \ref{joint_par} demonstrates the coupled binary partition, where the domains of two different sample sets are partitioned in tandem. The following property and theorem motivate us to construct co-BPM.

\begin{pro}
  For any pair of decision sequences $A_{l_1}$ and $A_{l_2}$ and their partitions $\{r_{i_1, l_1}\}_{i_1 = 1}^{l_1}$ and $\{r_{i_2, l_2}\}_{i_2 = 1}^{l_2}$, there exists a sequence $A_l$ and its partition $\{r_{i, l}\}_{i = 1}^l$ such that $r_{i_k, l_k}$ is the union of a subset of $\{r_{i, l}\}_{i = 1}^l$ for $k = 1, 2$ and $i_k = 1, ..., l_k$, namely, $A_l$ defines a finer partition than $A_{l_1}$ and $A_{l_2}$.
  \label{pro}
\end{pro}

The next theorem shows that a partition of $\Omega$ gives a lower bound of the discrepancies.
\begin{thm}
  Given a partition $\{r_{1, l}, ..., r_{l, l}\}$ of $\Omega$ at depth $l$, let \[\tilde{p}_i(x) = \sum_{k = 1}^l \frac{P_i(r_{k, l})}{|r_{k, l}|}\mathbf{1}\{x\in r_{k, l}\}\]
  where $i = 1, 2$ and $|\cdot|$ denotes the volume or size, then
  \begin{equation}
  D_\phi(p_1, p_2) \geq D_\phi(\tilde{p}_1, \tilde{p}_2)
  \label{lower_bound}
  \end{equation}
  namely, a partition gives a way to estimate $D_\phi$.
  \label{thm1}
\end{thm}
\begin{rem}
  Another lower bound \citep{Nguyen2010} of $D_\phi$ is $\sup_{f\in\mathcal{F}}\int [fdP_2 - \phi^*(f)dP_1]$, where $\mathcal{F}$ is a class of functions and $\phi^*$ is the conjugate dual function of $\phi$. Theorem \ref{thm1} shifts the difficulty of estimation from finding a good $\mathcal{F}$ to a good partition.
\end{rem}
\begin{rem}
   Under the condition that the integral $D_\phi(p_1, p_2)$ is Riemann integrable. It is trivial to show that $D_\phi(p_1, p_2) = \sup D_\phi(\tilde{p}_1, \tilde{p}_2)$, where the supremum is taken over all possible partitions at all depths.
  \label{rie}
\end{rem}
The proof is straightforward by applying the Jensen's inequality.
{\small
\begin{proof}
  We decompose the integral by $\{r_{1, l}, ..., r_{l, l}\}$ and apply Jensen's inequality,
  \begin{equation}
  \begin{split}
  &D_\phi(p_1, p_2) = \int_{\Omega} p_1(x)\phi\Big(\frac{p_2(x)}{p_1(x)}\Big)dx = \sum_{i = 1}^l\int_{r_{i, l}}p_1(x)\phi\Big(\frac{p_2(x)}{p_1(x)}\Big)dx\\
  &= \sum_{i = 1}^l P_1(r_{i, l})E_{p_1(\cdot|r_{i, l})}[\phi\Big(\frac{p_2(x)}{p_1(x)}\Big)] \geq \sum_{i = 1}^l P_1(r_{i, l})\phi\Big(E_{p_1(\cdot|r_{i, l})}\frac{p_2(x)}{p_1(x)}\Big) = \sum_{i = 1}^l P_1(r_{i, l})\phi\Big(\frac{P_2(r_{i, l})}{P_1(r_{i, l})}\Big)
  \end{split}
  \end{equation}
\end{proof}
}

Theorem \ref{thm1} provides to a lower bound of $f-$divergence on finite partitions, Table \ref{extreme} illustrates the importance of a good partition. Within each region, the gap between $D_\phi(p_1, p_2)$ and $D_\phi(\tilde{p}_1, \tilde{p}_2)$ comes from applying Jensen's inequality, therefore, it can be closed if $p_1(x)/p_2(x)$ in each region is approximately constant. Such observations indicate that finer partition of the domain would reduce the estimation bias. However, an overly fine partition will cause insufficient samples in each sub-region and inadvertently increase the overall estimation variance. So there is a \emph{trade-off between bias and variance}. Therefore, an appropriate partitioning respecting the tradeoff should reflect the landscape of $\mathcal{X}$ and $\mathcal{Y}$ simultaneously and avoid over-cutting. We will see how this intuition is implemented in our bayesian model.

\begin{table}
\begin{tabular}{cc}
\putindeepbox[10pt]{\includegraphics[width = .4\textwidth, height = .63in]{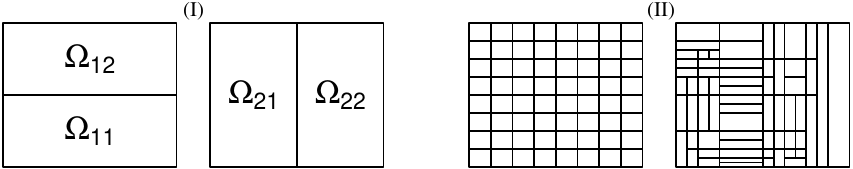}} &
    \putindeepbox[10pt]{\begin{tabular}{rcccc}
\hline
      & $D_{q = 1}$ & $D_{q = 2}$ & $D_{\alpha = 1}$ & $D_{\alpha = 2}$\\
      \hline
Hist & 0.5269 & 0.4935 & 0.7964 & 1.0291 \\
BP & \textbf{0.5431} & \textbf{0.4990} & \textbf{0.8267} & \textbf{1.0527} \\
Truth & 0.5518 & 0.5204 & 0.8604 & 1.0769 \\
\hline
\end{tabular}}
\end{tabular}
\caption{\textbf{A simple illustration demonstrates the importance of a good partition.} (I) $\mathcal{B}_1 = \{\Omega_{11}, \Omega_{12}\}$ and $\mathcal{B}_2 = \{\Omega_{21}, \Omega_{22}\}$ are two partitions of unit cube. Consider $p_1(x) = \frac{3}{2}\mathbf{1}\{x\in \Omega_{11}\} + \frac{1}{2}\mathbf{1}\{x\in \Omega_{12}\}$ and $p_2(x) = \frac{1}{2}\mathbf{1}\{x\in \Omega_{11}\} + \frac{3}{2}\mathbf{1}\{x\in \Omega_{12}\}$. Under $\mathcal{B}_1$, the lower bounds equal $D_\phi(p_1, p_2)$, i.e., $D_\phi = \frac{3}{4}\phi(\frac{1}{3}) + \frac{1}{4}\phi(3)$. For instance, the Total Variation Distance is $\frac{1}{2}$, Hellinger Distance is $\frac{\sqrt{3}}{2} - \frac{1}{2}$ and $\alpha-$divergence is $\alpha = 2$ is $\frac{1}{\alpha - 1}\log(\frac{3}{4}\cdot 3^{\alpha - 1} + \frac{1}{4}\cdot \frac{1}{3}^{\alpha - 1})$ respectively; however, under $\mathcal{B}_2$, the lower bounds are $\phi(1)$, where the Total Variation Distance, Hellinger Distance and $\alpha-$divergence with $\alpha = 2$ are all 0. (II) $p_1(x) = \beta_{3, 5}(x_1)\beta_{3, 5}(x_2)$ and $p_2(x) = \mathbf{1}\{x\in [0, 1]^2\}$. On the left is histogram by dividing each dimension into 8 equal sub-intervals; on the right is an adaptive partition with 64 sub-rectangles. As summarized in the right table, \textbf{BP} (adaptive binary partition) is closer to \textbf{Truth} (the true values) than \textbf{Hist} (histogram).}
  \label{extreme}
\end{table}

\section{Coupled Binary Partition Model}
In light of Property \ref{pro} and Theorem \ref{thm1}, for a given partition $\{r_{i, l}\}_{i = 1}^l$, we approximate the density $p_1$ and $p_2$ with $\hat{p}_1$ and $\hat{p}_2$ as
\normalsize\[\hat{p}_1(x) = \sum_{i = 1}^l \frac{m_{1i, l}}{|r_{i, l}|}\mathbf{1}\{x\in r_{i, l}\}, \hat{p}_2(y) = \sum_{i = 1}^l \frac{m_{2i, l}}{|r_{i, l}|}\mathbf{1}\{y\in r_{i, l}\}\]\normalsize
where $\sum_{i = 1}^l m_{1i, l} = 1$, $\sum_{i = 1}^l m_{2i, l} = 1$; $\mathbf{m}_{k, l} = (m_{k1, l}, ..., m_{kl, l})\geq \mathbf{0}, k = 1, 2$. Based on $\hat{p}_1$ and $\hat{p}_2$, we estimate divergence $D_{\phi}(p_1, p_2)$ as $D_{\phi}(\hat{p}_1, \hat{p}_2)$, where
\begin{equation}
D_{\phi}(\hat{p}_1, \hat{p}_2) = \sum_{i = 1}^l m_{1i, l}\phi(\frac{m_{2i, l}}{m_{1i, l}})
\label{div}
\end{equation}

The parameters $\mathbf{m}_{1, l}$, $\mathbf{m}_{2, l}$ and $\{r_{i, l}\}_{i = 1}^l$ are unknown, we assign a prior to them as follows: 1) Assume that $\{r_{i, l}\}_{i = 1}^l$ is generated by a decision sequence $A_l = (a_1, ..., a_{l - 1})$; 2) A decision sequence of depth $l\in\mathcal{N}^+$ has a prior density proportional to $\exp(-\sigma l)$ with some positive $\sigma$ and all decision sequences at the same depth are distributed uniformly---this part of the prior is to discourage over-cutting as discussed in the previous section; 3) $\mathbf{m}_{1, l}, \mathbf{m}_{2, l}$ have a Dirichlet prior $\textrm{Dir}(\delta, ..., \delta)$ respectively with some positive $\delta$ and the priors are independent between sample sets $\mathcal{X}$ and $\mathcal{Y}$. Thus,
 \normalsize\[p(l, A_l, (\mathbf{m}_{1, l}, \mathbf{m}_{2, l}))\propto \exp(-\sigma l)\prod_{i = 1}^l m_{1i, l}^{\delta - 1}\prod_{i = 1}^l m_{2i, l}^{\delta - 1}\]\normalsize

The likelihood of $\mathcal{X}$ and $\mathcal{Y}$ is
\normalsize\[p(\mathcal{X}, \mathcal{Y}|(l, A_l, (\mathbf{m}_{1, l}, \mathbf{m}_{2, l}))) = \prod_{i = 1}^l\Big(\frac{m_{1i, l}}{|r_{i, l}|}\Big)^{n_{1i, l}}\prod_{i = 1}^l\Big(\frac{m_{2i, l}}{|r_{i, l}|}\Big)^{n_{2i, l}}\]\normalsize

where $\mathbf{n}_k = (n_{ki, l})_{i = 1}^l, k = 1, 2$ denote the number of samples in each sub-region with respect to sample sets $\mathcal{X}$ and $\mathcal{Y}$. By denoting $\Theta = (l, A_l, (\mathbf{m}_{1, l}, \mathbf{m}_{2, l}))$, the posterior $\pi(\Theta|\mathcal{X}, \mathcal{Y})$ or $\pi(\Theta)$ for short is
\normalsize
\begin{equation}
\begin{split}
& \pi(\Theta = (l, A_l, (\mathbf{m}_{1, l}, \mathbf{m}_{2, l}))|\mathcal{X}, \mathcal{Y})\propto p(\Theta)p(\mathcal{X}, \mathcal{Y}|\Theta)\\
& \propto\exp(-\sigma l)\prod_{i = 1}^l\Big(\frac{1}{|r_{i, l}|}\Big)^{n_{1i, l} + n_{2i, l}}\prod_{i = 1}^l (m_{1i, l})^{\delta + n_{1i, l} - 1}\prod_{i = 1}^l (m_{2i, l})^{\delta + n_{2i, l} - 1}
\end{split}
\label{post}
\end{equation}
\normalsize

As discussed in Section 1, a naive two-step algorithm proceeds as follows: with $\mathcal{X}$ and $\mathcal{Y}$, estimating two piecewise constant densities $\hat{p}_1$ and $\hat{p}_2$ supported on (different) binary partitions independently; then computing $D_\phi(\hat{p}_1, \hat{p}_2)$, which requires to intersect all overlapped sub-regions between both partitions. The drawbacks of this type of approach are already discussed. However, according to Property \ref{pro}, any two piecewise constant densities supported on different binary partitions can be rewritten such that they are defined on the same binary partition, thus we \emph{couple} $\hat{p}_1, \hat{p}_2$ by partitioning their domain in tandem (i.e., forcing the same $A_l$) in the prior such that Theorem \ref{thm1} is applicable. The divergence of $\hat{p}_1, \hat{p}_2$ is estimated by $\mathbf{m}_{1, l}, \mathbf{m}_{2, l}$.

\subsection{Sampling}
Through sampling from the posterior, the divergence are estimated by $D_\phi(\hat{p}_1, \hat{p}_2)$. According to the decomposition in \eqref{post}, the depth $l$, the decision sequence $A_l$ and the probability masses $\mathbf{m}_{1, l}$, $\mathbf{m}_{2, l}$ can be sampled hierarchically. The first step is to generate the depth $l$ and decisions $A_l$. By marginalizing $\mathbf{m}_{1, l}$ and $\mathbf{m}_{2, l}$ in \eqref{post}, $(A_l, l)$ is distributed as
\normalsize
\begin{equation}
\pi(A_l, l|\mathcal{X}, \mathcal{Y})\propto \exp(-\sigma l)\beta((\delta + n_{1i, l})_{i = 1}^l)\beta((\delta + n_{2i, l})_{i = 1}^l)\prod_{i = 1}^l\Big(\frac{1}{|r_{i, l}|}\Big)^{n_{1i, l} + n_{2i, l}}
\label{post_par}
\end{equation}
\normalsize
where $\beta(\cdot)$ is the multinomial Beta function and defined as $\beta((z_i)_{i = 1}^k) = \prod_{i = 1}^k\Gamma(z_i) / \Gamma(\sum_{i = 1}^k z_i)$ and $\Gamma(z)$ is the Gamma function. Furthermore, conditioned on $l$, $A_l$ is distributed as
\normalsize
\begin{equation}
\pi(A_l|l, \mathcal{X}, \mathcal{Y})\propto \beta((\delta + n_{1i, l})_{i = 1}^l)\beta((\delta + n_{2i, l})_{i = 1}^l)\prod_{i = 1}^l\Big(\frac{1}{|r_{i, l}|}\Big)^{n_{1i, l} + n_{2i, l}}
\label{post_par}
\end{equation}
\normalsize

Once $A_l$ is generated, $\mathbf{m}_{1, l}$ and $\mathbf{m}_{2, l}$ are sampled through $\textrm{Dir}((\delta + n_{1i, l})_{i = 1}^l)$ and $\textrm{Dir}((\delta + n_{2i, l})_{i = 1}^l)$ respectively.

It is difficult to obtain the analytical distribution of $(A_l, l)$ because of the intractability of normalizing constant or partition function, Markov Chain Monte Carlo is employed to sample the posterior \eqref{post}. However, given the vast parameter space and the countless local modes, the naive Metropolis-Hastings \citep{hastings1970} suffers from slow mixing in our experience. In order to sample effectively, a proposal kernel should be equipped with the two properties: 1) it leverages the sequential structure of binary partition and the proposed partition reflects the difference amongst sub-regions (e.g., the sample counts of $\mathcal{X}$ and $\mathcal{Y}$) for rapid mixing; 2) the corresponding acceptance ratio depends on a smaller set of parameters, in other words, the transition probability is controlled by a subset of $\Theta$, such that the size (dimensionality) of searching space is reduced. Define
\normalsize
\begin{equation*}
g(\Theta' = (l', A_{l'}, (\mathbf{m}_{1, l'}, \mathbf{m}_{2, l'}))|\Theta = (l, A_l, (\mathbf{m}_{1, l}, \mathbf{m}_{2, l})))=p(A_{l'}, l'|A_l, l)p((\mathbf{m}_{1, l'}, \mathbf{m}_{2, l'})|A_{l'}, l)
\end{equation*}
\normalsize
where $p(A_{l'}, l'|A_l, l) = p(l'|l)p(A_{l'}|A_l, l')$ defines the jump probability, $l'$ is constrained to be $l - 1$ and $l + 1$, i.e., $p(l + 1|l) + p(l - 1|l) = 1$ and $p(l - 1|l) = 0$ when $l = 1$. In order to exploit the sequential structure of binary partition such that each $a_l$ is drawn with the guidance of $A_l$, $p(A_{l + 1}|A_l, l + 1) = p(a_l|A_l, l + 1)$ is defined as
\begin{equation}
\begin{split}
p(a_l|A_l, l + 1)& = \frac{\pi(A_{l + 1}|l + 1, \mathcal{X}, \mathcal{Y})}{\pi(A_{l}|l, \mathcal{X}, \mathcal{Y})}\\
&\propto\frac{\beta((\delta + n_{1i, l + 1})_{i = 1}^{l + 1})\beta((\delta + n_{2i, l + 1})_{i = 1}^{l + 1})}{\beta((\delta + n_{1i, l})_{i = 1}^l)\beta((\delta + n_{2i, l})_{i = 1}^l)}\frac{\prod_{i = 1}^l |r_{i, l}|^{n_{1i, l} + n_{2i, l}}}{\prod_{i = 1}^{l + 1}|r_{i, l + 1}|^{n_{1i, l + 1} + n_{2i, l + 1}}}
\label{individual}
\end{split}
\end{equation}
and $p(A_{l - 1}|A_l, l - 1) = p(l - 1|l)$ since $p(A_{l - 1}|A_l, l - 1) = 1$, $p((\mathbf{m}_{1, l'}, \mathbf{m}_{2, l'})|A_{l'}, l')$ is the joint Dirichlet distribution $\textrm{Dir}((\delta + n_{1i, l'})_{i = 1}^{l'})\times\textrm{Dir}((\delta + n_{2i, l'})_{i = 1}^{l'})$. Thus, the acceptance ratio is
\normalsize
\begin{equation}
\begin{split}
  &Q(\Theta\rightarrow\Theta') = \min\{1, \frac{\pi(\Theta')g(\Theta|\Theta')}{\pi(\Theta)g(\Theta'|\Theta)}\} = \\
  &\min\{1, \frac{\exp(\sigma l)\prod_{i = 1}^l|r_{i, l}|^{n_{1i, l} + n_{2i, l}}p(A_l, l|A_{l'}, l')\beta((\delta + n_{1i, l'})_{i = 1}^{l'})\beta((\delta + n_{2i, l'})_{i = 1}^{l'})}{\exp(\sigma l')\prod_{i = 1}^{l'}|r_{i, l'}|^{n_{1i, l'} + n_{2i, l'}}p(A_{l'}, l'|A_l, l)\beta((\delta + n_{1i, l})_{i = 1}^l)\beta((\delta + n_{2i, l})_{i = 1}^l)}\}
\end{split}
\label{acc_ratio}
\end{equation}
\normalsize
where $\pi(\Theta) = \pi(\Theta|\mathcal{X}, \mathcal{Y})$ is defined in \eqref{post}. According to \eqref{acc_ratio}, $\mathbf{m}_{1, l}, \mathbf{m}_{1, l'}$ and $\mathbf{m}_{2, l}, \mathbf{m}_{2, l'}$ are canceled, $Q(\Theta\rightarrow\Theta')$ only depends on $(A_l, l)$ and $(A_{l'}, l')$ which avoids searching for the vast space of $\mathbf{m}_{1, l}, \mathbf{m}_{2, l}$. Moreover, if $p(l'|l) = p(l|l')$ for $l > 1$, $Q(\Theta\rightarrow\Theta')$ can be further simplified.

In higher dimensions, sampling according to $p(A_{l'}|A_l, l')$ requires to count the number of points of $\mathcal{X}$, $\mathcal{Y}$ in each sub-region of $A_{l'}$, which utilizes the information of $\mathcal{X}, \mathcal{Y}$ but is expensive (with complexity $O((n_1 + n_2)d)$) in computation. If we are willing to run longer chains with cheaper cost per iteration, another heuristic choice for transition $p(A_{l'}, l'|A_l, l)$ is to keep $p(A_{l - 1}, l - 1|A_l, l) = p(l - 1|l)$ but take action $a_l$ uniformly: $p(A_{l + 1}, l + 1|A_l, l) = p(l + 1|l) / (l\cdot d)$ as there are $l\cdot d$ possible decisions (halving locations).
\begin{figure}
  \centering
  \includegraphics[width=.6\textwidth]{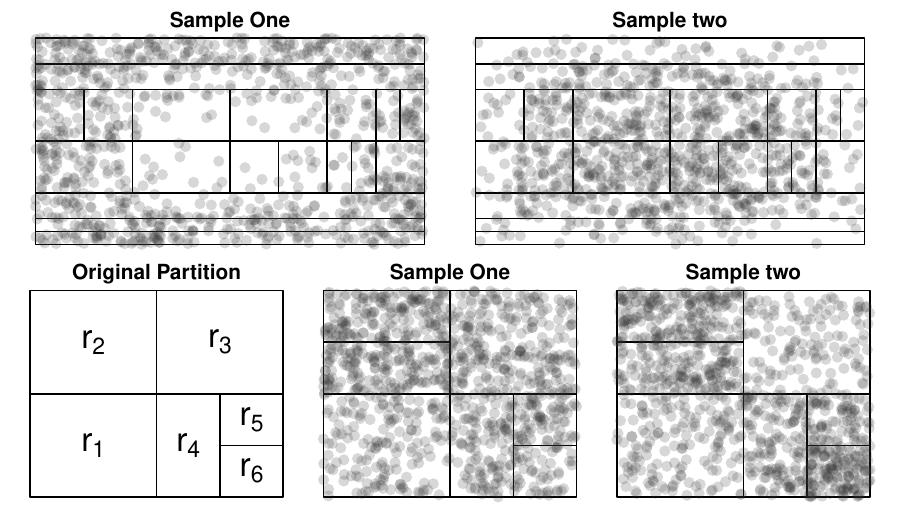}
  \caption{\textbf{Sanity Tests.} First row: 1,000 samples drawn from $p_{11}$ and $p_{21}$ and the learned partition; second row: 1,000 samples drawn from $p_{21}$ and $p_{22}$ and the learned partition.}
  \label{joint_sim}
\end{figure}
\section{Numeric Experiments}
\subsection{Sanity Tests for co-BPM}
We use two ``sanity tests'' to demonstrate that our model is sensitive to differences among samples, the densities used in experiments are two dimensional for ease of visualization. In the first group $p_{11}$ and $p_{12}$, we force the mixture of $p_{11}$ and $p_{12}$ is uniformly distributed, i.e., $\frac{1}{2}(p_{11} + p_{12}) = \mathbf{1}\{x\in [0, 1]^2\}$; we generate 1,000 points each, thus the combined sample is equivalent to 2,000 points from uniform distribution; as shown in the first row of Figure \ref{joint_sim}, their difference is revealed by the partition well. The second group $p_{21}$ and $p_{22}$ are chosen such that they are both supported on the binary partition and $p_{22}$ is defined on a finer partition than $p_{21}$. An effective model should be able to discover the underlying partition of $p_{22}$ or find a partition that is finer than both $p_{11}$ and $p_{22}$. The second row of Figure \ref{joint_sim} demonstrates such effectiveness of co-BPM.
\normalsize
\[p_{11}(x) = \frac{9}{5}\mathbf{1}\{x\in [0, 1]^2\} - \frac{4}{5}\beta_{2, 2}(x_1)\beta_{2, 2}(x_2)\]
 \[p_{12}(x) = \frac{1}{5}\mathbf{1}\{x\in [0, 1]^2\} + \frac{4}{5}\beta_{2, 2}(x_1)\beta_{2, 2}(x_2)\]
\[p_{21}(x) = \frac{2}{3}\mathbf{1}\{x\in r_1\} + \frac{4}{3}\mathbf{1}\{x\in r_2\} + \mathbf{1}\{x\in r_3\cup r_4\cup r_5\cup r_6\}\]
\[p_{22}(x) = \frac{2}{3}\mathbf{1}\{x\in r_1\} + \frac{4}{3}\mathbf{1}\{x\in r_2\} + \frac{1}{2}\mathbf{1}\{x\in r_3\} + \mathbf{1}\{x\in r_4\} + \frac{4}{3}\mathbf{1}\{x\in r_5\} + \frac{8}{3}\mathbf{1}\{x\in r_6\}\]
\normalsize

\subsection{Numeric Simulations}
\begin{figure}
\centering
  \includegraphics[width = .8\textwidth]{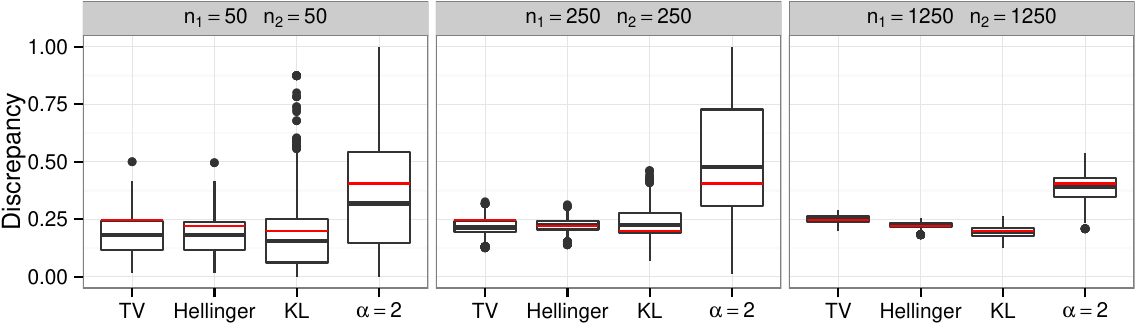}
  \caption{\textbf{1-dimensional simulation.} We draw 8,000 samples and generate box-plot after discarding the first 5,000 burn-in sample points. Theoretical divergences (red lines): 0.2461, 0.2207, 0.2000, 0.4056.}
  \label{one_dim}
\end{figure}

\begin{figure}
\centering
  \includegraphics[width = .8\textwidth]{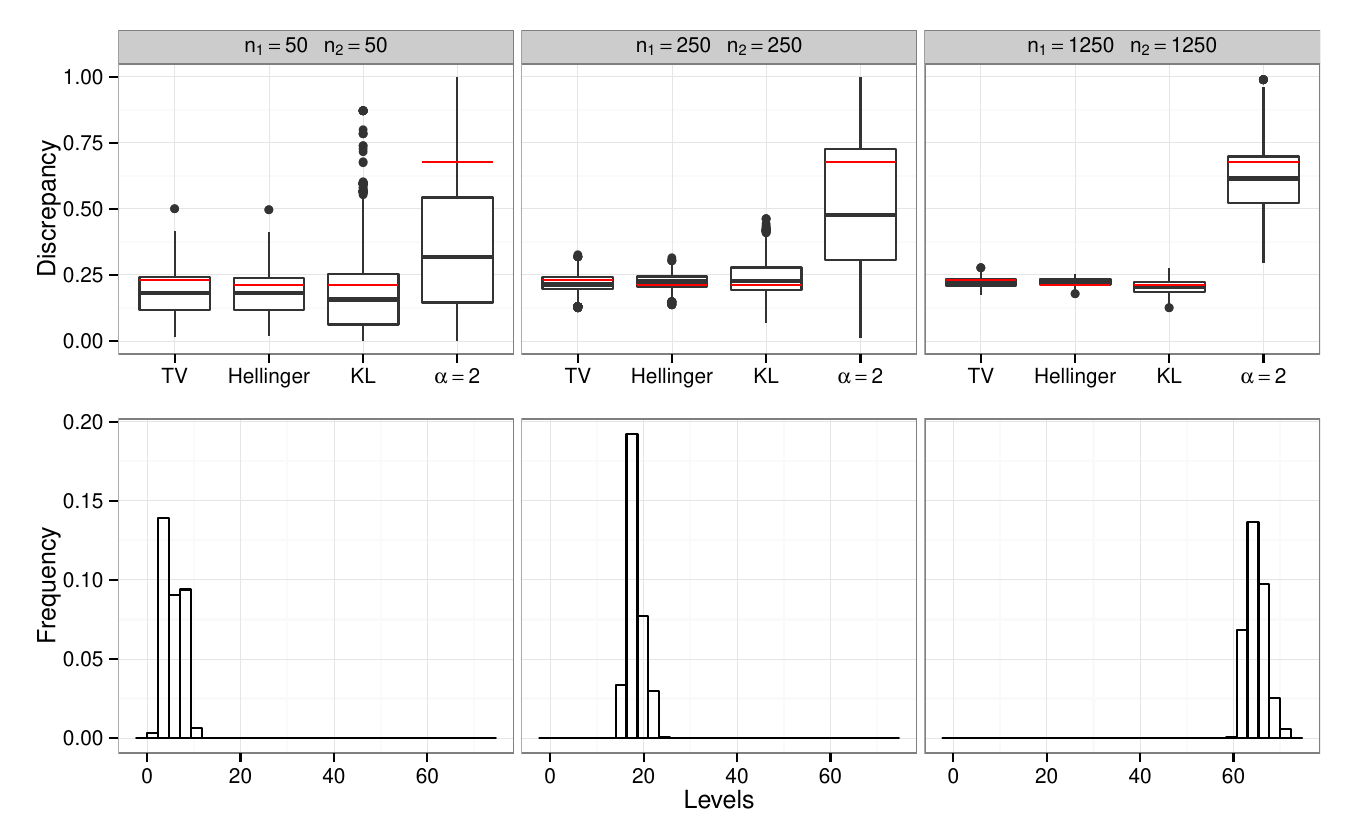}
  \caption{\textbf{3-dimensional simulation.} First row, box-plot with the same configuration as 1-dimensional case; theoretical divergences (red lines): 0.2301, 0.2129, 0.2133, 0.6769. Second row, we assess the behavior of depth $l$ after discarding first 5,000 burn-in samples, the increase of number of sub-regions as sample size increases indicates that co-BPM refines the partitions when more information becomes available, which is analogous to the multi-resolution property discussed in \citep{Wong2010}.}
  \label{three_dim}
\end{figure}

We demonstrate our methods by 2 simulations with dimension $d = 1, 3$ and size $n_1 = n_2 = 50, 250, 1250$. The parameters for co-BPM are $\delta = 1/2$ (which is the Jeffrey's non-informative prior), $\sigma = d + 1$, $p(l + 1|l) = p(l - 1|l) = 1/2$ for $l > 1$. The number of replicas for the box-plot is 3,000 and the burn-in number is 5,000.  The divergences we consider are Total Variation, Hellinger Distance and KL divergence and $\alpha-$divergence with $\alpha = 2$. The true values of the divergences are obtained by Monte Carlo method with $10^8$ samples.\\
\textbf{1-dimensional examples.} The densities are defined as below and the results are summarized in Figure \ref{one_dim}.
\normalsize\[p_1(x) = \beta_{6, 5}(x), p_2(x) = \beta_{5, 6}(x)\]\normalsize
where $\beta_{a, b}(x)$ is the Beta distribution with shape parameters $a, b$.
\\
\textbf{3-dimensional examples.} The densities are defined as below and the results are summarized in Figure \ref{three_dim}.
\begin{equation}
\begin{gathered}p_1(x, y, z) = \frac{2}{5}\beta_{1, 2}(x)\beta_{2, 3}(y)\beta_{3, 4}(z) + \frac{3}{5}\beta_{4, 3}(x)\beta_{3, 2}(y)\beta_{2, 1}(z) \\ p_2(x, y, z) = \frac{2}{5}\beta_{1, 3}(x)\beta_{3, 5}(y)\beta_{5, 7}(z) + \frac{3}{5}\beta_{7, 5}(x)\beta_{5, 3}(y)\beta_{3, 1}(z)
\end{gathered}
\label{3_example}
\end{equation}
\\
It is observed from Figure \ref{one_dim} and \ref{three_dim} that co-BPM estimates the divergences reasonably well. Moreover, the estimation errors and variances are decreased when the sample size increases, so is the number of outliers in box-plots. According the histogram of depth $l$ in Figure \ref{three_dim}, another interesting observation is that the number of sub-regions increases in tandem with the sample size, which indicates that co-BPM refines the partitions to reveal more structure of the sample sets as more information becomes available, this multi-resolution property of binary partition is also discussed in Optional P\'{o}lya Tree \citep{Wong2010}.

The vanishing boundaries of beta distribution cause the large range and variance of KL and $\alpha-$divergence where division is involved, as demonstrated by the previous examples. Instead of using Jeffrey's noninformative prior, we choose a stronger prior with larger $\delta$ as a tradeoff between bias with variance. According to law of total variance, $\mathrm{Var}(m_{1i}) = E[\mathrm{Var}(m_{1i}|\mathcal{X})] + \mathrm{Var}[E(m_{1i}|\mathcal{X})]$. Since $m_{1i}$ is generated from Dirichlet distribution, $\mathrm{Var}(m_{1i}|\mathcal{X}) = \frac{(n_{1i} + \delta)(n_1 + \delta l - n_{1i} - \delta)}{(n_1 + \delta l)^2(n_1 + \delta l + 1)} = O(\delta^{-1})$ and $\mathrm{Var}[E(m_{1i}|\mathcal{X})] = \mathrm{Var}(\frac{n_{1i} + \delta}{n_1 + \delta l}) = \frac{\mathrm{Var}(n_{1i})}{(n_1 + \delta l)^2}$; thus, asymptotically, its variance is reduced with a larger $\delta$, as shown in Table \ref{four_dim}.
\begin{table}
\center
\begin{tabular}{cc}
\putindeepbox[10pt]{\includegraphics[width=0.25\textwidth]{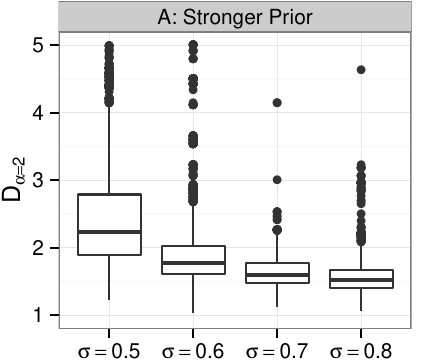}} &
    \putindeepbox[10pt]{\scalebox{1.3}{\begin{tabular}{@{}r|rrrr}
  \toprule
  & \multicolumn{4}{c}{Stronger Prior ($\delta$)}\\
  \hline
  & $0.5$ & $0.6$ & $0.7$ & $0.8$\\
  \midrule
  \hline
  Medians & 2.3057 & 1.7771 & 1.5960 & 1.5267\\
  Std & 1.2690 & 0.6643 & 0.2223  & 0.2494\\
  \bottomrule
\end{tabular}}}
\end{tabular}
\caption{\textbf{Illustration of bias and variance tradeoff.} $\mathcal{X}\sim\mathcal{N}(\mu_1\mathbf{1}, \sigma_1^2\mathbf{I})$, $\mathcal{Y}\sim\mathcal{N}(\mu_2\mathbf{1}, \sigma_2^2\mathbf{I})$ and are truncated in $[0, 1]^4$, where $|\mathcal{X}| = |\mathcal{Y}| = 500$; $\mu_1 = 1/3, \mu_2 = 1/2$ and $\sigma_1 = 1/5, \sigma_2 = 1/5$; $\mathbf{1}$ is 4-dim unit vector and $\mathbf{I}$ is 4-dim identity matrix. As pseudo-count $\sigma$ increases, the variance decreases and bias increases. True value: 2.2196. The table on the right lists the medians and standard errors.}
\label{four_dim}
\end{table}

\subsection{Comparison with Other Methods}
Most of other methods focus on KL divergence estimation. \citep{Nguyen2010} derive a general lower bound for $f-$divergence via conjugate dual function; however, all of their theories and experiments are based on KL divergence estimation in 1, 2, 3 dimensions and their algorithms require that two sample sets have same size. In this section, we compare co-BPM to the methods in \citep{Nguyen2010} and \citep{Perez-Cruz2008} in KL divergence estimation.

We briefly describe the methods in \citep{Nguyen2010} and \citep{Perez-Cruz2008}, interested readers may refer to the original papers for details. The core of \citep{Nguyen2010}'s algorithm is a convex program that has the number of parameters equal to the sample size. As pointed out by the authors, the performance of the algorithm depends on a regularization parameter $\lambda$; here we choose $\lambda$ according to their suggestions. There are two slightly different versions of KL estimator proposed, which are denoted by NWJ-M1 and NWJ-M2 in our experiments. On the other hand, the idea of \citep{Perez-Cruz2008}'s approach, as well as \citep{Perez-Cruz2008}'s, is relatively straightforward: the \emph{$k-$nearest neighbor (k-NN) density estimate} is computed for each sample in $\mathcal{X}$, i.e., $\hat{p}_{1}(x)$ and $\hat{p}_{2}(x)$ for $x\in\mathcal{X}$, then KL divergence is estimated by $\frac{1}{|\mathcal{X}|}\sum_{x\in\mathcal{X}}\log\frac{\hat{p}_{1}(x)}{\hat{p}_{2}(x)}$ (denoted by PC$-k$). The tuning parameter $k$ is critical to guarantee their performance, we follow the choice of \citep{Perez-Cruz2008} by setting $k = 1$ and $k = 10$. One should also notice that the positivity of the estimates is not guaranteed as $\hat{p}_1$ and $\hat{p}_2$ are not normalized densities.

\begin{figure}[!ht]
  \center
  \includegraphics[width = 0.8\textwidth]{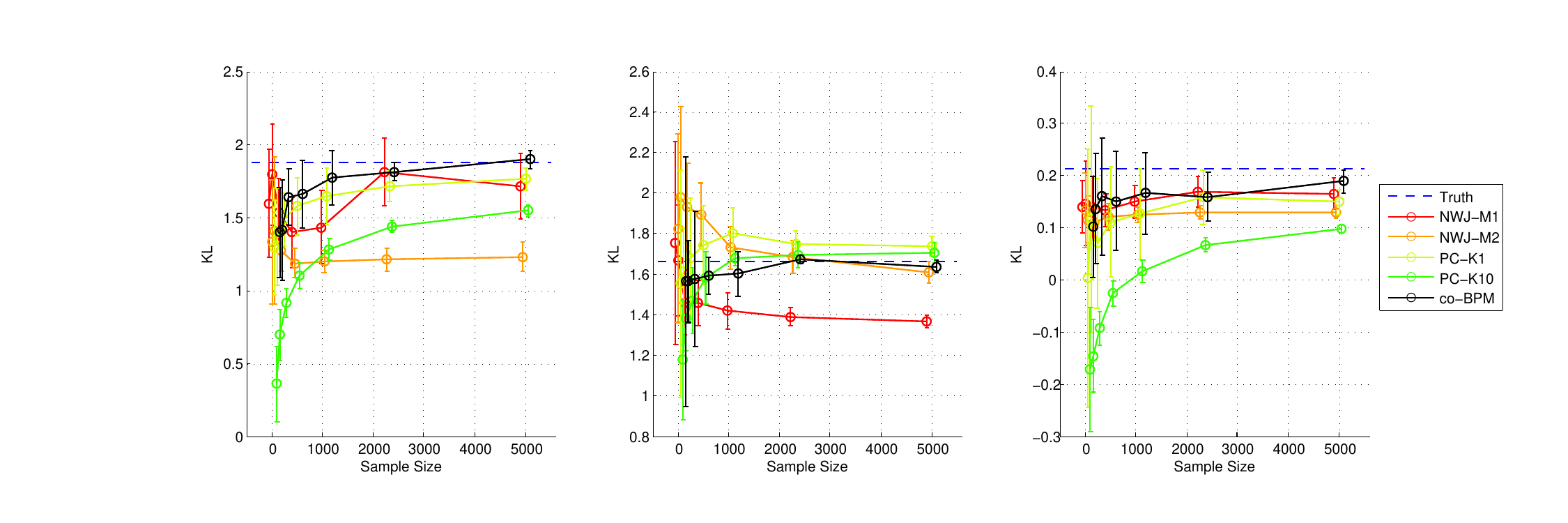}
  \caption{\textbf{Comparisons of NWJ, PC$-k$ and co-BPM.} All densities are truncated in $[0, 1]^3$ and each line is slightly shifted to avoid overlaps. Left: $p_1(x) \sim \mathbf{1}\{x\in [0, 1]^3\}$, $p_2(x) \sim \mathcal{N}(0, (1 / 3)^2\mathbf{I})$; Middle: $p_1(x) \sim \mathcal{N}(0, (1 / 2)^2\mathbf{I})$, $p_1(x) \sim \mathcal{N}(1, (1 / 2)^2\mathbf{I})$; Right: Beta mixture in \eqref{3_example}.}
  \label{comp}
\end{figure}

We report their performance in Figure \ref{comp}. For each of the three estimation problems described here, we experiment with 3-dim distributions and increase sample size from 50 to 5000 in logarithmic scale. Error bars are obtained by replicating each set-up 10 times. We see that co-BPM generally exhibits the best performance among the estimators considered. The estimates of NWJ-M1 and NWJ-M2 is somewhat less good and have larger variances; moreover, the convex program of NWJ-M1 or NWJ-M2 is computationally difficult because of the logarithm or entropy term and its scale increases rapidly with the sample size \citep{Grant2008}. However, PC$-k$ is more stable than NWJ and $k$ seems to strike a balance between bias and variance---the larger $k$ corresponds to lower variances but larger bias, e.g., its variance is smallest when $k = 10$ in Figure \ref{comp}. On the right plot, PC$-k$ produces negative estimates with small sample size as it does not guarantee positivity.
\subsection{Application to Image Clustering and Classification}\label{real_ex}
As a real data example, we apply co-BPM to image clustering and classification. The images are from \citep{Fei-Fei2005}. Each image is segmented to a collection of local patches, where each patch is a feature vector. Assuming that each patch is an independent draw from an underlying distribution, each image is represented by an iid sample set. Hence, the dissimilarity among images can be measured by the pairwise divergences of their corresponding sample sets.

We use the same setup as in \citep{Poczos2012}: 1) we randomly select 50 images from categories ``MITmountain'', ``MITcoast'', ``MIThighway'' and ``MITinsidecity'' respectively; 2) features are extracted as in \citep{Fei-Fei2005} and PCA is applied to reduce feature dimension to 2. The final training set has 200 sample sets (images) and each sample set contains 1600 2-dim points. We compare co-BPM with PC$-k$ using $\alpha-$divergence with $\alpha = 0.5$ and ``bag-of-words'' methods (BoW). BoW \citep{Bosch2006} quantizes each patch to 100 ``visual words'' and uses probability latent semantic analysis to convert each image to a low dimensional (20 in our experiment) probability vector called topic distributions. The pairwise dissimilarity of images are measured by Euclidean distances.

\begin{figure}[!ht]
  \center
  \includegraphics[width = 0.6\textwidth]{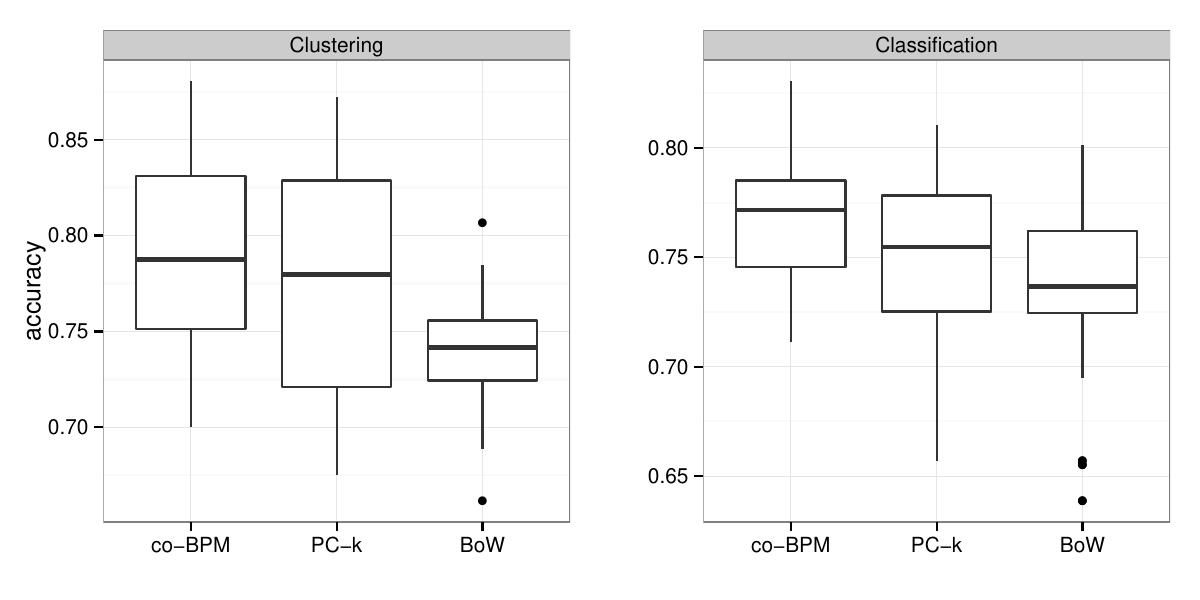}
  \caption{\textbf{Comparisons of NWJ, PC$-k$ and co-BPM.} All densities are truncated in $[0, 1]^3$ and each line is slightly shifted to avoid overlaps. Left: $p_1(x) \sim \mathbf{1}\{x\in [0, 1]^3\}$, $p_2(x) \sim \mathcal{N}(0, (1 / 3)^2\mathbf{I})$; Middle: $p_1(x) \sim \mathcal{N}(0, (1 / 2)^2\mathbf{I})$, $p_1(x) \sim \mathcal{N}(1, (1 / 2)^2\mathbf{I})$; Right: Beta mixture in \eqref{3_example}.}
  \label{real_data}
\end{figure}

With the dissimilarity matrices, we apply spectral clustering \citep{VonLuxburg2008} to cluster these images. In each cluster, we define cluster-wise accuracy as the percentage of the majority category. The overall accuracy is defined as the average of all cluster-wise accuracies. The procedure is repeated 20 times and the box-plot of the accuracies are reported in the left plot of Figure \ref{real_data}.

$\alpha-$divergence captures the difference between images pretty well. The median accuracies for co-BPM, PC$-k$ and BoW are 83.0\%, 74.9\% and 72.2\% respectively. The $t-$tests show that the difference between co-BPM and PC$-k$ ($p-$value $7.97\times 10^{-13}$) and between co-BPM and BoW ($p-$value $< 2.2\times 10^{-16}$) are significant.

Using the image labels and the pairwise divergences or distances, we can predict the labels via majority vote in $k-$NN classification. In each run, we draw additional 100 testing images randomly. We choose $k = 9$ and repeat 20 times, their classification accuracies are summarized in the right plot Figure \ref{real_data}.

Similar to the results as clustering, co-BPM (77.8\%) outperforms PC$-k$ and BoW by about 3\% and 5\% in median accuracy. The $t-$test of accuracies for (co-BPM, PC-$k$) and (co-BPM, BoW) have $p-$values $9.1\times 10^{-4}$ and $8.6\times 10^{-6}$.  One possible reason that co-BPM outperforms the other two is that co-BPM is adaptive enough to discover local discrepancies like $k-$NN but does not suffer from the normalization issue as PC$-k$.

\section{Conclusion and Discussion}
A unified single-shot approach to estimating divergences of distributions is proposed from a Bayesian perspective. The experiments demonstrate its attractive empirical performance. In applications, this approach can be naturally extended to handle multiple-sample-set cases, i.e., we partition multiple sample sets jointly and estimate their densities as piecewise constant function on the same binary partition. Then their pairwise divergences can be compuated efficiently by \eqref{div}, which is useful in clustering and classification as in Section \ref{real_ex}. Another direction for future work is to understand the theoretical properties of co-BPM, such as the asymptotic behavior as well as the convergence rate.
\bibliographystyle{apalike}
  \bibliography{nips2014}
\end{document}